\documentclass[12pt,preprint]{aastex}
\usepackage{emulateapj5}

\def\gta{\ifmmode {\mathbin{\lower 3pt\hbox   
    {$\,\rlap{\raise 5pt\hbox{$\char'076$}}\mathchar"7218\,$}}}
    \else {${\mathbin{\lower 3pt\hbox
    {$\rlap{\raise 5pt\hbox{$\char'076$}}\mathchar"7218\,$}}}
    $}\fi}
\def\lta{\ifmmode {\,\mathbin{\lower 3pt\hbox   
    {$\,\rlap{\raise 5pt\hbox{$\char'074$}}\mathchar"7218\,$}}}
    \else {${\mathbin{\lower 3pt\hbox
    {$\rlap{\raise 5pt\hbox{$\char'074$}}\mathchar"7218\,$}}}
    $}\fi}

\shorttitle {Flame Spreading on Neutron Stars}
\shortauthors {Bhattacharyya and Strohmayer}

\begin{document}

\title {Thermonuclear Flame Spreading on Rapidly Spinning Neutron Stars:
Indications of the Coriolis Force?}

\author {Sudip Bhattacharyya\altaffilmark{1,2}, and Tod
E. Strohmayer\altaffilmark{3}}

\altaffiltext{1}{CRESST and X-ray Astrophysics Lab, Astrophysics
Science Division, NASA's Goddard Space Flight Center, Greenbelt, MD
20771; sudip@milkyway.gsfc.nasa.gov} \altaffiltext{2}{Department of
Astronomy, University of Maryland, College Park, MD 20742}
\altaffiltext{3}{X-ray Astrophysics Lab, Astrophysics Science
Division, NASA's Goddard Space Flight Center, Greenbelt, MD 20771;
stroh@clarence.gsfc.nasa.gov}

\begin{abstract}

Millisecond period brightness oscillations during the intensity rise
of thermonuclear X-ray bursts are likely caused by an azimuthally
asymmetric, expanding burning region on the stellar surface. The time
evolution of the oscillation amplitude during the intensity rise
encodes information on how the thermonuclear flames spread across the
stellar surface.  This process depends on properties of the accreted
burning layer, surface fluid motions, and the surface magnetic field
structure, and thus can provide insight into these stellar
properties. We present two examples of bursts from different sources
that show a decrease in oscillation amplitude during the intensity
rise. Using theoretical modeling, we demonstrate that the observed
amplitude evolution of these bursts is not well described by a
uniformly expanding circular burning region.  We further show that by
including in our model the salient aspects of the Coriolis force (as
described by Spitkovsky, Levin, and Ushomirsky) we can qualitatively
reproduce the observed evolution curves. Our modeling shows that the
evolutionary structure of burst oscillation amplitude is sensitive to
the nature of flame spreading, while the actual amplitude values can
be very useful to constrain some source parameters.

\end{abstract}

\keywords{relativity --- stars: neutron --- stars: rotation ---
  X-rays: binaries --- X-rays: bursts --- X-rays: individual (4U
  1636--536, SAX J1808.4--3658)}

\section {Introduction} \label{sec: 1}

Millisecond period brightness oscillations, ``burst oscillations'',
during thermonuclear X-ray bursts (Strohmayer \& Bildsten 2006;
Woosley \& Taam 1976; Lamb \& Lamb 1978) from accreting neutron stars
were first discovered with the Rossi X-ray Timing Explorer ({\it
RXTE}) in 1996 (Strohmayer et al.). Since then, a total of 19 low mass
X-ray binaries (LMXBs; including three tentative detections; Thompson
et al. 2005; Kaaret et al. 2007; Bhattacharyya 2007) have exhibited
this timing feature (see Lamb \& Boutloukos 2007; Bhattacharyya et
al. 2006; Markwardt et al. 2007). The discovery of burst oscillations
at the spin frequencies of the accreting millisecond pulsars SAX
J1808.4--3658 and XTE J1814--338 (see Chakrabarty et al. 2003;
Strohmayer et al. 2003) has conclusively linked these oscillations to
neutron star spin, and showed that the burst oscillation frequency
gives a direct measure of the stellar spin frequency. This discovery
has also established that burst oscillations originate at the stellar
surface, and hence can be very useful to constrain neutron star mass
and radius (Miller, \& Lamb 1998; Muno et al. 2002).  Measurements of
these stellar parameters are critical to understand the nature of
high-density matter inside neutron stars, which is a fundamental
problem of physics (Bhattacharyya et al. 2005).

Thermonuclear X-ray bursts also provides a powerful tool to probe the
physics of flame spreading under the extreme conditions that exist on
neutron stars. This is because bursts ignite at a particular point on
the stellar surface, and then spread to burn all the surface fuel
(Fryxell \& Woosley 1982; Spitkovsky et al. 2002; Bhattacharyya \&
Strohmayer 2006a; 2006b). As flame spreading depends on shearing flows
in the surface layers (Spitkovsky et al. 2002), the study of this
spreading can be very useful to understand surface fluid motions.
Moreover, flame spreading can be helpful for mapping the magnetic
field structure on the stellar surface. This is because the shearing
flows present during flame spreading can make this field locally
strong, which may in turn act back on the flow (Spitkovsky et
al. 2002; Bhattacharyya \& Strohmayer 2006c).  However, a detailed
theoretical study of flame spreading including all relevant physical
effects has not yet been done, perhaps partly due to a lack of
observational motivation. Such motivation includes the significant
detection and measurement of the diagnostic features, such as the
evolution of spectral and burst oscillation properties during the
spreading.  But, flame spreading is expected to occur during burst
rise, and in this short period (sub-second to a few seconds) the
significant detection of the diagnostic aspects is difficult due to
the statistical quality of the present data.

Recently Spitkovsky et al. (2002) have proposed a theoretical model of
flame spreading on rapidly spinning neutron stars including the
effects of the Coriolis force. According to these authors, after
ignition the Coriolis force is initially not important due to the
small size of the burning region (which implies a large Rossby
number). During this time the flame spreads quickly with the so-called
geostrophic speed. So, the burning region (hot spot) attains a
considerable size almost at the beginning of the burst rise.  After
this, for a rapidly spinning neutron star the Coriolis force becomes
important, and the flame spreads as a shearing flow with the much
slower ageostrophic speed.  This speed decreses as the burning front
latitude increases, which causes a portion of the burning region to
encircle the stellar equator quickly, maintaining an asymmetric
portion at higher latitudes for some more time (for mid-latitude
ignition; Fig. 8c of Spitkovsky et al. 2002; see also Bhattacharyya \&
Strohmayer 2006c).

Strohmayer et al. (1997) found that during the rise of some bursts
from the LMXB 4U 1728--34, the oscillation amplitude diminished, which
is expected for an expanding burning region during burst rise.  These
initial results suggested that the tracking of amplitude evolution can
be a useful probe of flame spreading. However, more detailed
observational measurements of amplitude evolution have been found and
reported only recently (for 4U 1636--536 and SAX J1808.8--3658;
Bhattacharyya \& Strohmayer 2005; 2006c). In this Letter (in \S~2), we
first show that these more detailed evolutionary curves exhibit
amplitude changes (decreasing with burst rise). Then, using
theoretical models we demonstrate that these observed curves cannot be
easily explained with a uniformly expanding hot spot. Finally, we show
that models that include the salient features of the Coriolis force
can qualitatively reproduce the observed curves. In \S~3, we discuss
the implications of our analysis.

\section {Analysis} \label{sec: 2}

Tracking the burst oscillation amplitude evolution during burst rise
requires a strong oscillation throughout the rise. Therefore, given
the (1) small duty cycle of bursts, (2) non-detection of burst rise
oscillations for some bursts, (3) short burst rise period, (4) low
count rate during the early burst rise, and the fact that only {\it
RXTE} can currently detect burst oscillations, so far clear amplitude
evolution curves (for burst rise) have been reported only for two
bursts (See Fig. 1) from two rapidly spinning neutron stars (4U
1636--536: spin frequency $\nu_{\rm *} = 582$ Hz; and SAX
J1808.8--3658: $\nu_{\rm *} = 401$ Hz). Hence, if flame spreading is
influenced by the Coriolis force, then the bursts from these two
sources should show this influence. The evolutionary curve for each of
these bursts is consistent with a rapid amplitude decay, followed by a
nearly constant level of significantly non-zero oscillation amplitude
for some time. Before using these observed properties to probe flame
spreading, first we show that the data are not consistent with a
constant amplitude level. In order to demonstrate this, we have fitted
each data set with a constant amplitude, and the best-fit levels are
shown in Fig. 1. The resulting $\chi^2/\nu$ for these levels (20.7/7
for 4U 1636--536, and 18.1/9 for SAX J1808.8--3658) clearly shows the
significant decrease of the amplitude during burst rise. Motivated by
this, we have theoretically computed oscillation amplitude evolution
for an expanding hot spot on a neutron star's surface assuming various
source parameter values. Although some previous studies explored such
model computations (Strohmayer et al. 1997; Nath et al. 2002;
Strohmayer 2004), here we calculate and compare various models in
detail for the first time.

We have started with the simplest spreading geometry, viz, a uniformly
expanding circular hot spot. If the effect of stellar spin (and hence
the Coriolis force) on spreading is negligible, then this may be a
reasonable approximation to how flames actually spread. This is
because, with a relatively low stellar magnetic field ($10^7-10^9$ G)
for LMXBs, the accreted matter (fuel) should not be confined, and
should be present all over the surface almost uniformly.  In our
calculations, we have considered eight source parameters (see also
Bhattacharyya et al. 2005): (1) the dimensionless neutron star
radius-to-mass ratio $R/M$, (2) the stellar mass $M$, (3) $\nu_{\rm
*}$, (4) the observer's inclination angle $i$, (5) the polar angle
$\theta_{\rm c}$ of the hot spot center, (6) the angular radius
$\Delta \theta$ of the spot, (7) the blackbody temperature of the spot
$T_{\rm BB}$ (burst spectra can normally be well fitted with a
blackbody model; Strohmayer \& Bildsten 2006), and (8) a parameter $n$
that gives a measure of the beaming in the emitter's frame, where the
specific intensity as a function of the angle $\psi$ (in the emitter's
frame) from the surface normal is $I(\psi) \propto \cos^n\psi$. This
beaming may be due to the scattering of burst photons in an optically
thick layer of thermal electrons.  Note that for semi-infinite
plane-parallel layers with a constant net flux and Thomson scattering,
$n \sim 0.5$ (Chandrasekhar 1960).  In our calculation of energy
dependent flux from a hot spot, we have taken into account (1) Doppler
and special relativistic effects, (2) gravitational redshift, and (3)
light-bending (in a Schwarzschild spacetime). In order to include
light-bending, we have backtracked the paths of the photons from the
observer to the source (see Bhattacharyya et al. 2001; 2005).  The
model light curve is calculated by repeating the same procedure for
many identical spots at different $\phi$-positions (but the same
$\theta$-position) on the surface of the star.  The actual phase
points of the light curve are calculated from these $\phi$-positions,
the stellar spin frequency, and the time delay considerations. The
time delays result from the fact that photons emitted at different
points on the stellar surface take different times to reach the
observer. We have then folded this model light curve with a suitable
{\it RXTE} PCA response matrix, and determined the oscillation
amplitude in the same way it was determined from the data.  We have
repeated this procedure for many hot spots with increasing $\Delta
\theta$ (for a chosen combination of other parameter values) to
determine the oscillation amplitude evolution.

In Fig. 2, we have plotted these evolutionary curves for various
parameter values. Note that, here, we only show the points with
fractional peak amplitude less than 1.  Fig. 2 clearly shows that the
model curves for widely different parameter values are qualitatively
very similar, that is, the amplitude first decreases slowly, and then
decreases rapidly. On the other hand, the observed curves of Fig. 1
first decrease rapidly, and then maintain a significantly non-zero
level. A qualitative comparison of Fig. 1 \& 2 suggests that the
uniform expansion of a circular hot spot cannot easily explain the
observed amplitude evolution during rise for these bursts.  Fig. 2
also shows that the oscillation amplitude increases with the increase
of $R/M$, $M$, $\nu_{\rm *}$ $i$, $\theta_{\rm c}$ (except when
$\theta_{\rm c}$ is close to $180^{\rm o}$), and $n$, and with the
decrease of $\Delta \theta$ and $T_{\rm BB}$. Among these parameters,
$i$, $\theta_{\rm c}$, $R/M$ and $n$ affect the amplitude the most
(apart from $\Delta \theta$, which is not a free parameter for an
evolutionary curve), with the first two having the maximum effects.
Consequently, these parameter values can be measured more confidently
from oscillation amplitudes (see the last sentence of \S~2).  We note
that a large variation of $T_{\rm BB}$ does not change the amplitude
very much, which justifies (given the quality of data) our use of the
same blackbody temperature throughout the flame spreading.

Motivated by the suggestion that the flames probably do not spread
isotropically and uniformly on a rapidly spinning neutron star's
surface, we included some qualitative features of the Coriolis force
(mentioned in \S~1) in our modeling.  For a chosen combination of
source parameter values, we started with an already large hot spot
(because of initially fast geostrophic flow; \S~1) in a chosen
$\phi$-range and $\theta$-range in the low- to mid-latitudes. Most of
the hot spot was in the northern hemisphere (through which the
observer's line of sight passes), and crossed the equator only
slightly. Then we allowed a very narrow $\theta$-portion (of the spot)
on the equator to spread in the $\phi$-directions with a speed
proportional to $1/\sqrt{\cos\theta}$ (Spitkovsky et al. 2002).  Here,
we assumed an average $\theta$-value, and kept the $\phi$-direction
speed unchanged until a $\phi$-symmetric belt around the equator was
formed. This narrow $\phi$-expanding portion was also allowed to
simultaneously expand in $\theta$-directions with a speed proportinal
to $1/\sqrt{\cos\theta}$, where $\theta$ is the instantaneous polar
angle of the northern and southern burning fronts.  In addition, the
north edge of the original hot spot was allowed to expand northwards
with the same speed formula.  As a result of such spreading, the hot
spot had two portions (see Fig. 3): (1) an equatorial portion that
encircled the equator quickly, and expanded north and southwards
simultaneously (but progressively at a slower rate); and (2) a
$\phi$-asymmetric portion, that moved northwards, and became gradually
narrower (in $\theta$-direction) because of the assumed spreading
speed formula.  The rapid formation of the equatorial belt is expected
to cause a fast decay of the amplitude, and subsequently the mid- to
high-latitude $\phi$-asymmetric portion should give rise to a lower
amplitude that decreases (more slowly) primarily because of the
increasing persistent contribution from the expanding $\phi$-symmetric
belt. Our model calculation shows this expected evolution (see
Fig. 4), which qualitatively remains the same for widely different
values of $R/M$, $M$, $\nu_{\rm *}$ $i$, $T_{\rm BB}$, and $n$, and is
qualitatively similar to the observed evolution shown in Fig. 1. Here
we note that, while the models show the amplitude evolution from the
beginning of flame spreading (except the initial very rapid spreading)
up to the later part, the data of Fig. 1 may be for an intermediate
time phase of spreading. This is because, the observed oscillation
power may not be significant (1) in the initial phase due to the low
count rate, and (2) in the final phase due to the low
amplitude. Finally, although the evolutionary structures of the models
of Fig. 2 and the models of Fig. 4 are different, the dependence of
amplitude on each source parameter is the same for both sets of
models.

\section {Discussion and Conclusions} \label{sec: 3}

In this Letter, we have, for the first time, reported a detailed
comparison among theoretical models of burst oscillation amplitude
evolution during burst rise in order to probe the flame spreading
phenomenon.  Our fitting of burst rise data from two different bursts
from two rapidly spinning neutron stars indicate that the oscillation
amplitude first decays quickly, and then maintains a non-zero
near-constant level.  We have shown that such behavior cannot be
qualitatively explained with a simple uniform expansion of a circular
hot spot.  However, we note that if the circular hot spot first
expands rapidly, and then, after attaining a large size (covering most
of the stellar surface), expands with more than an order of magnitude
slower speed, then the observed amplitude curves may be
reproduced. But, such a dramatic time dependence of expansion may not
be realistic. Inclusion of Coriolis force effects, on the other hand,
can provide a natural explanation of the observed amplitude curves.
We have included some salient features of these effects (based on the
work of Spitkovsky et al. 2002) into our spreading model, and found
that these new model amplitude evolutions are qualitatively similar to
those observed.  This suggests that we may be seeing, for the first
time, effects of the Coriolis force on the surface layers of fast
spinning neutron stars.  This also provides additional motivation to
theoretically study thermonuclear flame spreading phenomena
considering all the main physical effects.  Our modeling also shows
that the time evolution of burst oscillation amplitude is sensitive to
the nature of flame spreading, while the actual amplitude values
strongly depend on some source parameters, such as $i$, $R/M$,
etc. Therefore, understanding flame spreading, and constraining the
source parameters should be achieveable with detailed modeling of
higher signal to noise ratio observations of the burst oscillation
sources.  Such data may be obtainable with future X-ray missions such
as {\it ASTROSAT}, {\it Constellation-X}, {\it XEUS}, or a large-area
``Super-{\it RXTE}'' timing mission.

\acknowledgments

{}

\clearpage
\begin{figure}
\vspace{-7.0 cm}
\hspace{-2.5 cm}
\epsscale{0.8}
\plotone{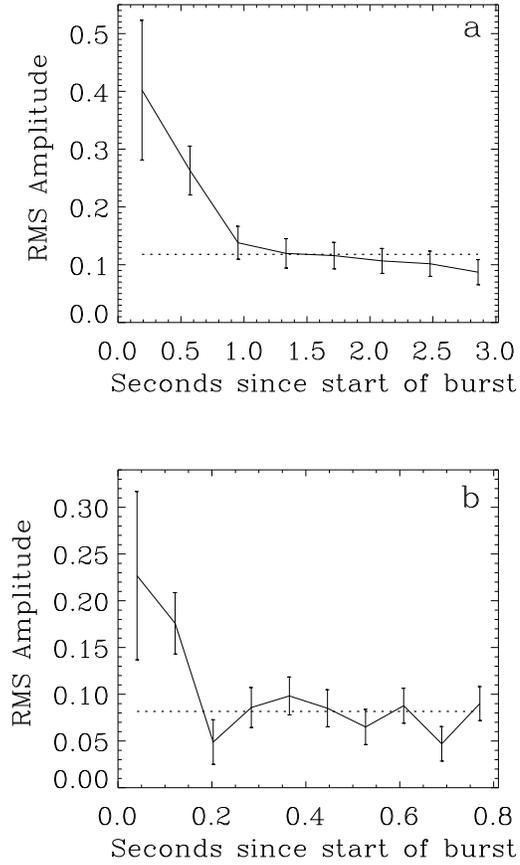}
\vspace{-3.0 cm}
\caption {Observed evolution of burst oscillation amplitude during the
rising portions of two thermonuclear X-ray bursts (solid lines with
error bars). Panel {\it a}: for a burst from the LMXB 4U 1636--536
(Fig. 1 of Bhattacharyya \& Strohmayer 2005); and panel {\it b}: for a
burst from the LMXB SAX J1808.4--3658 (Fig. 1 of Bhattacharyya \&
Strohmayer 2006c). The dotted lines give the best-fit constant
amplitude levels, and show that the amplitudes evolve significantly.
}
\end{figure}

\clearpage
\begin{figure}
\vspace{-3.0 cm}
\hspace{-1.9 cm}
\epsscale{0.5}
\plotone{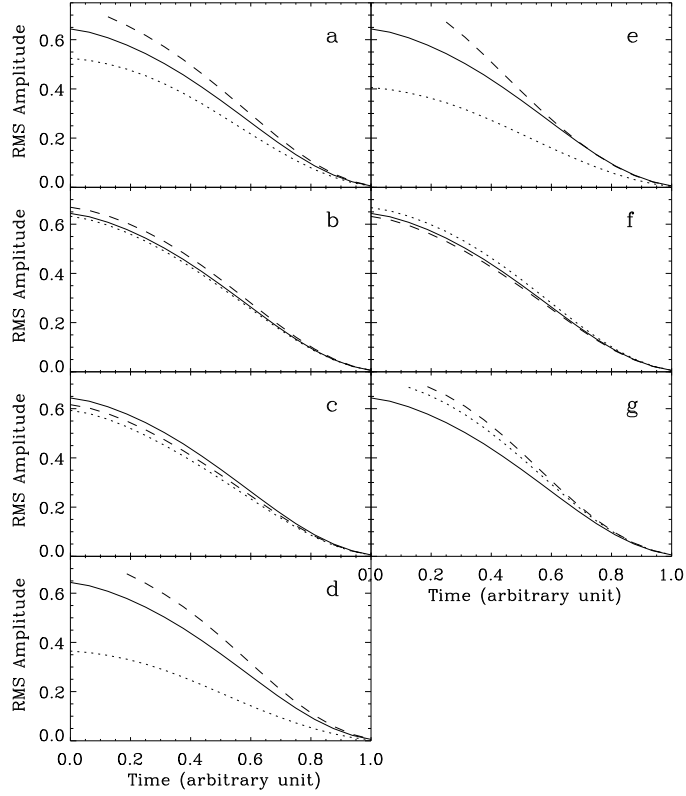}
\vspace{0.5 cm}
\caption {Theoretical models of burst oscillation amplitude evolution
for a uniformly expanding circular hot spot. For each panel, the solid
curve is for the parameter values (see \S~2): $R/M = 4.5$, $M = 1.6
M_{\odot}$, $\nu_{\rm *} = 582$ Hz, $i = 60^{\rm o}$, $\theta_{\rm c}
= 70^{\rm o}$, $T_{\rm BB} = 1.7$ K, and $n = 0$. In each panel, we
change one parameter value, while keeping the other parameter values
fixed at those for the solid curve. Panel {\it a}: dotted curve: $R/M
= 3.8$, dashed curve: $R/M = 5.5$; panel {\it b}: dotted: $M = 1.4
M_{\odot}$, dashed: $M = 2.0 M_{\odot}$; panel {\it c}: dotted:
$\nu_{\rm *} = 200$ Hz, dashed: $\nu_{\rm *} = 401$ Hz; panel {\it d}:
dotted: $i = 30^{\rm o}$, dashed: $i = 80^{\rm o}$; panel {\it e}:
dotted: $\theta_{\rm c} = 40^{\rm o}$, dashed: $\theta_{\rm c} =
130^{\rm o}$; panel {\it f}: dotted: $T_{\rm BB} = 1.0$ K, dashed:
$T_{\rm BB} = 2.5$ K; and panel {\it g}: dotted: $n = 0.5$, dashed: $n
= 0.8$. This figure shows that the qualitative nature of burst
oscillation amplitude evolution for a uniformly expanding circular hot
spot for any source parameter values is qualitatively different from
those observed (see Fig. 1).  }
\end{figure}

\clearpage
\begin{figure}
\hspace{-1.9 cm}
\epsscale{0.5}
\plotone{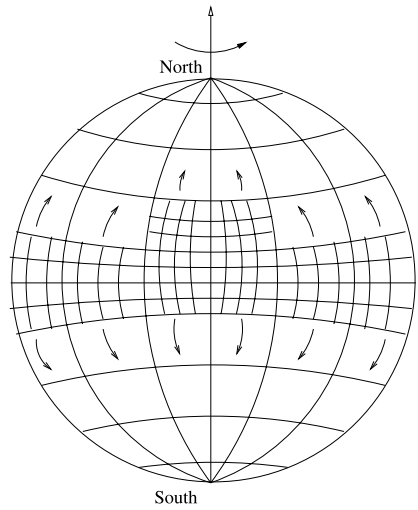}
\vspace{-0 cm}
\caption {Schematic diagram of flame spreading on a spinning neutron
star including salient features of the Coriolis force. We have used
this spreading scheme to compute our model evolutionary curves in
Fig. 4.  The spin axis and the direction of spin are shown. The meshed
region represents the hot spot, and the direction and speed of
expansion are indicated by the arrows.  The flame speed is higher at
lower latitudes, as approximately shown by the the length of the
arrows.  Here, the equatorial belt has already ignited, and the
residual $\phi$-asymmetry in the northern hemisphere gives rise to the
low amplitude oscillations.  }
\end{figure}

\clearpage
\begin{figure}
\hspace{-1.9 cm}
\epsscale{0.5}
\plotone{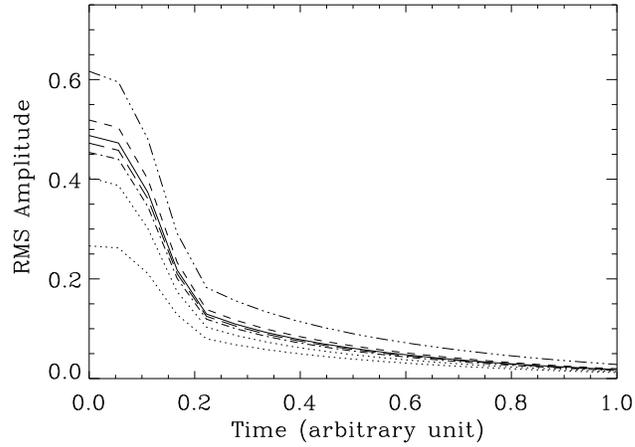}
\vspace{1.0 cm}
\caption {Theoretical models of burst oscillation amplitude evolution
for a hot spot expanding in a way that captures some features of the
Coriolis force (see \S~2; also see Fig. 3). The solid curve is for the
parameter values (see \S~2): $R/M = 4.5$, $M = 1.6 M_{\odot}$,
$\nu_{\rm *} = 582$ Hz, $i = 60^{\rm o}$, $T_{\rm BB} = 1.7$ K, and $n
= 0$. For each of the other curves, we change one parameter value,
while keeping the other parameter values fixed at those for the solid
curve. Upper dotted curve: $R/M = 3.8$; short-dashed curve: $M = 2.0
M_{\odot}$; dash-dot curve: $\nu_{\rm *} = 401$ Hz; lower dotted
curve: $i = 30^{\rm o}$; long-dashed curve: $T_{\rm BB} = 2.5$ K; and
dash-triple-dot curve: $n = 0.8$. This figure shows that the
qualitative nature of amplitude evolution for this model is similar to
those observed (see Fig. 1).  }
\end{figure}

\end{document}